\documentclass[seceq,preprint]{jpsj2}

\def\runauthor{
Kazumasa {\sc Miyake}
}

\title{
Effect of Spin-Orbit Interaction in Spin-Triplet Superconductor: 
Structure of ${\bf d}$-vector and Anomalous $^{17}$O-NQR Relaxation 
in Sr$_2$RuO$_4$
}

\author{
Kazumasa {\sc Miyake} 
}

\inst
{
Division of Materials Physics, 
Department of Materials Engineering Science, 
Graduate School of Engineering Science, 
Osaka University, Toyonaka, Osaka 560-8531, Japan
}
\recdate
{
October 21, 2009
}

\abst
{
Supposing the spin-triplet superconducting state of Sr$_2$RuO$_4$, 
the spin-orbit (SO) coupling associated with relative motion in Cooper pairs 
is calculated by extending the method for the dipole-dipole 
coupling given by Leggett in the superfluid $^{3}$He.  
It is shown that the SO coupling 
works only in the equal-spin pairing (ESP) state to make the pair angular 
momentum $\hbar{\vec L}$ and the pair spin angular momentum 
${\rm i}{\vec d}\times{\vec d}^{*}$ parallel with each other.  
The SO coupling gives rise to the internal Josephson effect in a chiral 
ESP state as in superfluid A-phase of $^3$He with a help of an additional 
anisotropy arising from SO coupling of atomic origin which works to direct the 
{\bf d}-vector into $ab$-plane.  This resolves the problem of the 
anomalous relaxation of $^{17}$O-NQR and the structure of {\bf d}-vector 
in Sr$_2$RuO$_4$.
}

\kword
{spin-orbit coupling, triplet superconductivity, Sr$_2$RuO$_4$, 
NQR relaxation rate
}

\begin{document}
\sloppy
\maketitle
\section{Introduction}
Nature of superconductivity of Sr$_2$RuO$_4$ has attracted much attention 
since its discovery by Maeno and his coworkers in 1994~\cite{Maeno1}.  
Now it seems to have been 
accepted that the Cooper pair is in the spin-triplet state~\cite{Ishida1}.  
However, its gap structure of the triplet state has not been confirmed yet.  
Especially, the intrinsic direction of the {\bf d}-vector is still under 
debate.

The effect of spin-orbit (SO) interaction {\it associated with relative 
motion of the Cooper pair}, together with an atomic SO interaction 
and the Hund's rule coupling, is expected to be crucial for determining the 
stable superconducting state among possible states in the manifold of the triplet 
pairing states.  It will be shown that effects of SO 
interaction are much more crucial than that of the magnetic dipole-dipole (DD) 
interaction which was essential for clarifying the nature of superfluid phase 
of liquid $^3$He~\cite{Leggett}.  

A highlight at earlier stages of research of Sr$_2$RuO$_4$ is a result of 
NMR Knight shift by 
Ishida {\it et al}\/\cite{Ishida1} which exhibits no decrease across 
the transition temperature $T_{\rm c}$ when the magnetic field {\bf B} of the 
order of Tesla is in the $ab$ (RuO$_2$) plane, suggesting that the Copper pairing 
is in the triplet state and the {\bf d}-vector is in the direction parallel to 
the $c$-axis, 
perpendicular to the plane.  
This structure of {\bf d}-vector was predicted by a group theoretical argument 
on the assumption that orbital and spin space are transformed together due to the 
``strong" SO coupling in forming the Cooper pairs~\cite{Rice}.  
However, the assumption of the ``strong" SO coupling is not self-evident.  

From experimental side, we should have been much more 
careful to draw a conclusion about the gap structure from this fact because 
the {\bf d}-vector has a tendency to rotate in such a way that {\bf d} and 
{\bf B} are perpendicular with each other even within the $ab$-plane.  
Without the magnetic filed {\bf B} or under the sufficiently low field, 
the direction of {\bf d}-vector is determined 
by other small perturbation such as, DD or SO interaction (two-body and/or 
single-body), or the effect of 
sample boundary.  The magnetic field of $\sim$Tesla seems too large to draw 
the conclusion about an intrinsic nature of the gap, considering the low 
condensation energy of $\sim k_{\rm B}T_{\rm c}$ with $T_{\rm c}\simeq 1.5$K.  
Concerning this subtlety, we should remember the case of UPt$_3$, 
in which the Knight shift shows no decrease for all 
direction of the magnetic field $B>$0.5 Tesla~\cite{Tou1}, while 
it shows clear decrease across $T_{\rm c}$ when $B<$0.2 Tesla is applied 
along $b$- or $c$-direction, suggesting the intrinsic direction of {\bf d} 
is in the $bc$-plane~\cite{Tou2,Yotsuhashi}.  

Indeed, six years later, Murakawa {\it et al}\/ reported that the Knight shift 
does not exhibit the decrease across $T_{\rm c}$ also in the case where the small 
field {\bf B} ($0.02<B<0.05$ Tesla) is along the $c$-axis~\cite{Ishida2,Ishida3}.  
This implies that the direction of ${\bf d}$-vector identified by the former 
experiment is not intrinsic but is forced by the magnetic field.  
The most plausible interpretation of these Knight shift measurements made 
down to the low magnetic field of $B\sim 0.02$Tesla is that the intrinsic 
direction of the {\bf d}-vector is in the $ab$-plane and the anisotropy 
field in the $ab$-plane is smaller than 0.05Tesla at most.  The latter 
conclusion is derived from the fact that the Knight shift does not exhibit 
any decrease across the $T_{\rm c}$ down to the magnetic field of 
$B=0.05$Tesla perpendicular to the $c$-axis~\cite{Ishida3}. 
Therefore, theories justifying the {\it fact} that {\bf d} is parallel to the 
$c$-axis might lose their plausibility, and other explanations are anticipated.  
Quite recently, it was shown that the intrinsic direction of {\bf d}-vector 
can be in the $ab$-plane if the Coulomb interaction among electrons on the 
2p orbitals of O (other than that on the 4d orbital of Ru) is taken into 
account\cite{Hoshihara} together with the atomic SO interaction and the Hund's 
rule coupling at Ru site~\cite{Yoshioka}. 

A role of the SO interaction between orbital and 
spin angular momentum of Cooper pairs may be crucial because it works to 
make the ${\bf d}$-vector perpendicular to {\bf L}, the pair angular 
momentum in the $c$-axis; therefore, the {\bf d}-vector is in the $ab$-plane.  
This effect may open a way to resolve another puzzle of anomalous NQR 
relaxation of $^{17}$O in the superconducting regime~\cite{Mukuda}.  
Indeed, the analysis shows that the dynamical spin susceptibility 
$\sum_{\bf q}{\rm Im} \chi_{zz}({\bf q},\omega)/\omega|_{\omega=\omega_{\rm NQR}}$ 
exhibits a huge enhancement at $0<T<T_{\rm c}$ while 
$\sum_{\bf q}{\rm Im} \chi_{xx}({\bf q},\omega)/\omega|_{\omega=\omega_{\rm NQR}}$ and 
$\sum_{\bf q}{\rm Im} \chi_{yy}({\bf q},\omega)/\omega|_{\omega=\omega_{\rm NQR}}$ follow the $T^{3}$-law at $0<T<T_{\rm c}$, the canonical $T$-dependence for the 
anisotropic superconductivity with a gap with line(s) of zero~\cite{Mukuda,Ishida4}.  
A possible explanation for this behavior is that 
the internal Josephson effect arises through the SO coupling and 
gives rise to the excess NQR relaxation other than the conventional 
relaxation in the superconducting state due to the quasiparticles contribution.  
The purpose of this paper is to develop a theoretical framework to explain the 
effect of SO interaction of Cooper pairs in the equal-spin-pairing (ESP) state, and 
to resolve the puzzle of anomalous NQR relaxation observed in Sr$_2$RuO$_4$.

\section{Spin-Orbit Coupling of Cooper Pairs in Triplet State}
The SO interaction associated with the relative motion of quasiparticles 
is given as follows:
\begin{equation}
H_{\rm so}=-{\mu_{\rm B}^{2}\over \hbar}{m_{\rm band}\over m^{*}}
\sum_{i}\sum_{j\not=i}
{1\over r_{ij}^{3}}{\vec{\sigma}}_{i}\cdot
\left[{\vec r}_{ij}\times[(2{\bar g}-1){\vec p}_{i}
-2{\bar g}{\vec p}_{j}]\right],
\label{spinorbit1}
\end{equation}
where $\mu_{\rm B}$ is the Bohr magneton, $m_{\rm band}$ band mass of electron, 
$m^{*}$ effective mass of quasiparticles, and ${\bar g}$ is defined 
as ${\bar g}\equiv\mu_{\rm eff}/\mu_{\rm B}$, $\mu_{\rm eff}$ being 
the effective magnetic moment.  The difference of prefactor of 
${\vec p}_i$ from that of ${\vec p}_j$ is due to the so-called Thomas 
precession by the effect of the special relativity.  
The appearance of the factor $m_{\rm band}/m^{*}$ in (\ref{spinorbit1}) 
is due to the vertex correction for angular momentum, $m_{\rm band}/m^{*}a$, 
derived on an extended Ward-Pitaevskii identity as shown in Appendix~\cite{AGD}, 
and that for the spin density, $1/a$~\cite{AGD,Leggett1964}.  The renormalization 
amplitude ``$a$" in the denominators is cancelled by the weight of quasiparticles ``$a$".  

In this paper, we restrict our discussions within the triplet manifold of 
ESP.  By the procedure similar to that described in Ref.\ \cite{Leggett} for 
the dipole interaction, the interaction (\ref{spinorbit1}) leads to 
the SO free energy $F_{\rm so}$,  for the Cooper pairs in the 
chiral state with the pair angular momentum $\hbar{\vec L}$, as follows:
\begin{equation}
F_{\rm so}=-g_{\rm so}({\rm i}{\vec d}\times{\vec d}^{*})\cdot{\vec L},
\label{spinorbit2}
\end{equation}
where the coefficient $g_{\rm so}$ depends on the details of the dispersion 
of the quasiparticles and the pairing interaction.  
In the spherical model of three dimensions (3d), $g_{\rm so}$ is given as 
\begin{equation}
g_{\rm so}=g_{\rm d}{ m_{\rm band}\over m^{*}}\times 4
{4{\bar g}-1\over {\bar g}^{2}},
\label{spinorbit3}
\end{equation}
where $g_{\rm d}$ is the strength of the dipolar coupling in the 
``ESP"-superconducting state in 3d.  In the cylindrical model or in two 
dimensions (2d), $g_{\rm so}$ is given as 
\begin{equation}
g_{\rm so}=g_{\rm d}{ m_{\rm band}\over m^{*}}\times 8
{4{\bar g}-1\over {\bar g}^{2}}.
\label{spinorbit3a}
\end{equation}

Hereafter, we derive expressions, (\ref{spinorbit2}) and 
(\ref{spinorbit3}) or (\ref{spinorbit3a}), starting with 
the interaction Hamiltonian (\ref{spinorbit1}) which is represented 
in the second quantization as follows: 
\begin{eqnarray}
H_{\rm so}&=&-{\mu_{\rm B}^{2}\over \hbar}{ m_{\rm band}\over m^{*}}
\int\int{\rm d}{\bf r}_{1}{\rm d}{\bf r}_{2}
{1\over |{\bf r}_{1}-{\bf r}_{2}|^{3}}
\psi_{\alpha}^{\dagger}({\bf r}_{1})\psi_{\gamma}^{\dagger}({\bf r}_{2})
{\vec \sigma}_{\alpha\beta}\delta_{\gamma\delta}
\nonumber
\\
& &\qquad\qquad\cdot
\left[({\vec r}_{1}-{\vec r}_{2})\times(-{\rm i}\hbar)
\left((2{\bar g}-1){\vec{\nabla}}_{1}-
2{\bar g}{\vec{\nabla}}_{2})\right)\right]
\psi_{\delta}({\bf r}_{2})\psi_{\beta}({\bf r}_{1}).
\label{so3}
\end{eqnarray}
By introducing the relative coordinate ${\bf r}\equiv{\bf r}_{1}-{\bf r}_{2}$, 
and the center of mass coordinate ${\bf R}\equiv({\bf r}_{1}+{\bf r}_{2})/2$, 
(\ref{so3}) is reduced to 
\begin{eqnarray}
H_{\rm so}&=&-{\mu_{\rm B}^{2}\over \hbar}{ m_{\rm band}\over m^{*}}
\int\int{\rm d}{\bf R}{\rm d}{\bf r}
{1\over r^{3}}
\psi_{\alpha}^{\dagger}({\bf R}+{\bf r}/2)
\psi_{\gamma}^{\dagger}({\bf R}-{\bf r}/2)
{\vec \sigma}_{\alpha\beta}\delta_{\gamma\delta}
\nonumber
\\
& &\qquad\qquad\qquad
\cdot\left[{\vec r}\times(-{\rm i}\hbar)
\left((4{\bar g}-1){\vec{\nabla}}_{r}-{1\over 2}{\vec{\nabla}}_{R}
\right)\right]
\psi_{\delta}({\bf R}-{\bf r}/2)\psi_{\beta}({\bf R}+{\bf r}/2).
\label{so4}
\end{eqnarray}

The free energy due to the SO coupling is given by the expectation value 
of $H_{\rm so}$, (\ref{so4}).  As in the case of dipole-dipole 
interaction~\cite{Leggett}, we rely on the following decoupling 
approximation: 
\begin{eqnarray}
& &
\langle\psi_{\alpha}^{\dagger}({\bf R}+{\bf r}/2)
\psi_{\gamma}^{\dagger}({\bf R}-{\bf r}/2)
\psi_{\delta}({\bf R}-{\bf r}/2)\psi_{\beta}({\bf R}+{\bf r}/2)\rangle
\nonumber
\\
& &
\qquad\qquad
\simeq\langle\psi_{\alpha}^{\dagger}({\bf R}+{\bf r}/2)
\psi_{\gamma}^{\dagger}({\bf R}-{\bf r}/2)\rangle
\langle\psi_{\delta}({\bf R}-{\bf r}/2)\psi_{\beta}({\bf R}+{\bf r}/2)\rangle
\nonumber
\\
& &
\qquad\qquad
=\langle\psi_{\alpha}^{\dagger}({\bf r}/2)
\psi_{\gamma}^{\dagger}(-{\bf r}/2)\rangle
\langle\psi_{\delta}(-{\bf r}/2)\psi_{\beta}({\bf r}/2)\rangle.
\label{so5}
\end{eqnarray}
Then, since (\ref{so5}) is independent of {\bf R}, 
\begin{equation}
F_{\rm so}\equiv
\langle H_{\rm so}\rangle=-{\mu_{\rm B}^{2}\over \hbar}{ m_{\rm band}\over m^{*}}
({4\bar g}-1)V
\int{\rm d}{\bf r}{1\over r^{3}}
{\vec \sigma}_{\alpha\beta}\delta_{\gamma\delta}
\cdot 
F_{\gamma\alpha}^{*}({\bf r})[{\vec r}\times(-{\rm i}\hbar)
{\vec{\nabla}}_{r})]F_{\delta\beta}({\bf r}),
\label{so6}
\end{equation}
where $V$ is the system volume and 
\begin{equation}
F_{\delta\beta}({\bf r})\equiv
\langle\psi_{\delta}({\bf r}/2)\psi_{\beta}(-{\bf r}/2)\rangle.
\label{so7}
\end{equation}
In terms of the conventional notation,
\begin{equation}
F_{\alpha\beta}({\bf r})={\rm i}({\vec \sigma}\sigma_{2})_{\alpha\beta}
\cdot {\vec F}({\bf r}),
\label{so8}
\end{equation}
(\ref{so6}) is expressed as 
\begin{eqnarray}
F_{\rm so}&=&-{\mu_{\rm B}^{2}\over \hbar}{ m_{\rm band}\over m^{*}}
({4\bar g}-1)V
\int{\rm d}{\bf r}{1\over r^{3}}
{\sigma}^{i}_{\alpha\beta}\delta_{\gamma\delta}
(\sigma_{k}\sigma_{2})^{*}_{\beta\gamma}
(\sigma_{\ell}\sigma_{2})_{\beta\delta}
\nonumber
\\
& & \qquad\qquad\qquad\qquad\qquad\qquad\qquad\qquad
F_{k}^{*}({\vec r})[{\vec r}\times(-{\rm i}\hbar)
{\vec{\nabla}}_{r})]_{i}F_{\ell}({\bf r}).
\label{so9}
\end{eqnarray}
With the use of the identity, 
\begin{equation}
\sigma_{\alpha\beta}^{i}\delta_{\gamma\delta}
(\sigma_{k}\sigma_{2})^{*}_{\beta\gamma}
(\sigma_{\ell}\sigma_{2})_{\beta\delta}=
{\rm Tr}(\sigma_{2}\sigma_{k}\sigma_{i}\sigma_{\ell}\sigma_{2})
=2{\rm i}\epsilon_{i\ell k},
\label{so10}
\end{equation}
(\ref{so9}) is reduced to 
\begin{equation}
F_{\rm so}=-{\mu_{\rm B}^{2}\over \hbar}{ m_{\rm band}\over m^{*}}
2{\rm i}\epsilon_{i\ell k}({4\bar g}-1)V
\int{\rm d}{\bf r}{1\over r^{3}}F_{k}^{*}({\bf r})
[{\vec r}\times(-{\rm i}\hbar){\vec{\nabla}}_{r})]_{i}
F_{\ell}({\bf r}).
\label{so11}
\end{equation}
Since the vector pairing amplitude $F_{\ell}$ is the eigen 
function of the relative angular momentum, the following 
relation holds: 
\begin{equation}
[{\vec r}\times(-{\rm i}\hbar){\vec{\nabla}}_{r})]_{i}
F_{\ell}({\bf r})=\hbar L_{i}F_{\ell}({\bf r}).
\label{so12}
\end{equation}
Then, with the use of the conventional definition of the {\bf d}-vector as 
\begin{equation}
{\vec F}({\bf r})\equiv F({\bf r}){\vec d},
\label{so13}
\end{equation}
(\ref{so9}) is expressed as 
\begin{equation}
F_{\rm so}=-{\mu_{\rm B}^{2}\over \hbar}{ m_{\rm band}\over m^{*}}
({4\bar g}-1)2{\rm i}\epsilon_{i\ell k}d^{*}_{k}d_{\ell}\hbar L_{i}
V\int{\rm d}{\bf r}{1\over r^{3}}|F({\bf r})|^{2}.
\label{so14}
\end{equation}
Thus, the free energy due to the SO coupling is given as 
the following form:
\begin{equation}
F_{\rm so}=-g_{\rm so}
({\rm i}{\vec d}\times{\vec d}^{*})\cdot{\vec L},
\label{so15}
\end{equation}
where the coupling constant $g_{\rm so}$ is expressed as 
\begin{equation}
g_{\rm so}=\mu_{\rm B}^{2}{ m_{\rm band}\over m^{*}}
({4\bar g}-1)2V\int{\rm d}{\bf r}{1\over r^{3}}|F({\bf r})|^{2},
\label{so16}
\end{equation}
where the pair amplitude $F({\bf r})$ is given by its {\bf k}-representation as 
\begin{equation}
F({\bf r})=\sum_{k}e^{{\rm i}{\bf k}\cdot{\bf r}}F({\bf k}).
\end{equation}

Explicit form of $F({\bf r})$ depends on the type of pairing and 
dimensionality of space.  Let us first examine the ABM state in 3d case: 
Then the pairing amplitude $F({\bf r})$ is given as~\cite{Leggett} 
\begin{equation}
F({\vec r})=\Psi\int{{\rm d}{\hat {\bf k}}\over 4\pi}
\sqrt{{3\over 2}}({\hat k}_{x}+{\rm i}{\hat k}_{y})
e^{{\rm i}k_{\rm F}({\bf k}_{\rm F}\cdot{\bf r})}.
\label{so17}
\end{equation}
The ${\hat {\bf k}}$-integration of the part including ${\hat k}_{x}$ is 
performed as 
\begin{equation}
\int{{\rm d}{\hat {\bf k}}\over 4\pi}
{\hat k}_{x}e^{{\rm i}k_{\rm F}({\bf k}_{\rm F}\cdot{\bf r})}
=-{\rm i}{x\over r}
\left[{\cos k_{\rm F}r\over k_{\rm F}r}-
{\sin k_{\rm F}r\over (k_{\rm F}r)^{2}}\right].
\label{so18}
\end{equation}
The integration of the part including ${\hat k}_{y}$ is performed 
similarly, leading to the expression for the pairing amplitude $F({\bf r})$ 
as follows:
\begin{equation}
F({\bf r})=\sqrt{{3\over 2}}\Psi
{-{\rm i}x+y\over r}\left[{\cos k_{\rm F}r\over k_{\rm F}r}-
{\sin k_{\rm F}r\over (k_{\rm F}r)^{2}}\right].
\label{so18a}
\end{equation}
Then, substituting (\ref{so18a}), the {\bf r}-integration in (\ref{so16}) 
is performed as 
\begin{eqnarray}
\int{\rm d}{\bf r}{1\over r^{3}}|F({\bf r})|^{2}&=&
4\pi\Psi^{2}\int_{0}^{\infty}{{\rm d}r\over r}
\left[{\cos k_{\rm F}r\over k_{\rm F}r}-
{\sin k_{\rm F}r\over (k_{\rm F}r)^{2}}\right]^{2}
\nonumber
\\
&=&\pi\Psi^{2},
\end{eqnarray}
where we used the formula of definite integral
\begin{equation}
\int_{0}^{\infty}{{\rm d}t\over t}\left[{{\cos t}\over t}-
{{\sin t}\over t^{2}}\right]={1\over 4}
\label{so19}
\end{equation}
As a result, the coupling constant $g_{so}$, (\ref{so16}), 
is expressed as follows:
\begin{equation}
g_{\rm so}=\mu_{\rm B}^{2}{ m_{\rm band}\over m^{*}}
({4\bar g}-1)2\pi\Psi^{2}V
\label{so20}
\end{equation}

This result should be compared with that for the dipole-dipole interaction 
given by Leggett for the ABM state:~\cite{Leggett}
\begin{equation}
g_{\rm d}={\pi\over 2}\mu_{\rm eff}^{2}\Psi^{2}
={\pi\over 2}{\bar g}^{2}\mu_{\rm B}^{2}\Psi^{2}.  
\label{so21}
\end{equation}
Therefore, the relation (\ref{spinorbit3}) holds.  
With the use of this coupling $g_{\rm d}$, the free energy due to the 
dipole-dipole interaction is given as\cite{Leggett}  
\begin{equation}
F_{\rm d}=-{3\over 5}g_{\rm d}
|({\vec d}\cdot{\vec L})|^{2}.
\label{so22}
\end{equation}

In the case of ABM state in 2d, the pair amplitude 
$F({\bf r})=F({\vec \rho})$ is given as 
\begin{equation}
F({\vec \rho})=\Psi\int{{\rm d}{\hat {\bf k}}\over 2\pi}
({\hat k}_{x}+{\rm i}{\hat k}_{y})
e^{{\rm i}k_{\rm F}({\bf k}_{\rm F}\cdot{\vec \rho})},
\label{so23}
\end{equation}
where ${\vec \rho}=(x,y)$ is 2d vector in $xy$-plane.  
The ${\hat {\bf k}}$-integration for the first term in (\ref{so23}) is 
performed as follows:
\begin{eqnarray}
\int{{\rm d}{\hat {\bf k}}\over 2\pi}
{\hat k}_{x}e^{{\rm i}k_{\rm F}({\bf k}_{\rm F}\cdot{\vec \rho})}
& &=-{{\rm i}\over k_{\rm F}}{\partial\over \partial x}
\left(\int{{\rm d}{\hat {\bf k}}\over 2\pi}e^{{\rm i}k_{\rm F}({\bf k}_{\rm F}
\cdot{\vec \rho})}\right)
\nonumber
\\
& &=-{{\rm i}\over k_{\rm F}}{\partial J_{0}(k_{\rm F}\rho)\over \partial x}
\nonumber
\\
& &=-{\rm i}{x\over \rho}J_{1}(k_{\rm F}\rho),
\label{so24}
\end{eqnarray}
where $J_{n}(z)$ is the Bessel function of the $n$-th order.  By a similar 
calculation for the second term in (\ref{so23}), the pair amplitude 
(\ref{so23}) is given as 
\begin{equation}
F({\vec \rho})=\Psi{-{\rm i}x+y\over \rho}J_{1}(k_{\rm F}\rho)
\label{so25}
\end{equation}

Then, substituting (\ref{so25}), the {\bf r}-integration in (\ref{so16}) 
is performed as follows:
\begin{eqnarray}
\int{\rm d}{\bf r}{1\over r^{3}}|F({\bf r})|^{2}&=&
\int{\rm d}{\vec \rho}|F({\vec \rho})|^{2}
\int_{-\infty}^{\infty}{\rm d}z{1\over(\rho^{2}+z^{2})^{3/2}}
\nonumber
\\
&=&2\int{\rm d}{\vec \rho}{1\over \rho^{2}}|F({\vec \rho})|^{2}
\nonumber
\\
&=&4\pi\Psi^{2}\int_{0}^{\infty}
{{\rm d}\rho \over \rho}[J_{1}(k_{\rm F}\rho)]^{2}
\nonumber
\\
&=&2\pi\Psi^{2},
\label{so26}
\end{eqnarray}
where we have used the formula of definite integral
\begin{equation}
\int_{0}^{\infty}{{\rm d}t\over t}[J_{1}(t)]^{2}={1\over 2}.
\label{so27}
\end{equation}
Therefore, the coupling constant $g_{\rm so}$, (\ref{so16}), is expressed as 
\begin{equation}
g_{\rm so}=\mu_{\rm B}^{2}{ m_{\rm band}\over m^{*}}
({4\bar g}-1)4\pi\Psi^{2}V,
\label{so28}
\end{equation}
leading to the relation (\ref{spinorbit3a}).  

In the case of 2d system Sr$_2$RuO$_4$, the free energy of dipole-dipole 
interaction 
is given as follows\cite{Hasegawa}:
\begin{equation}
F_{\rm d}=-{3c\over 4a\pi}g_{\rm d}
\left[({\vec d}\cdot{\vec L})^{2}-{1\over 3}\right],
\label{so28a}
\end{equation}
where $a$\/(=3.87\AA) and $c$\/(=6.37\AA) are the lattice constant of 
the primitive cell of Sr$_2$RuO$_4$~\cite{Mack}.  

\section{Non-Unitary State due to Spin-Orbit Coupling}
The SO coupling (\ref{so15}) induces the non-unitary component of 
{\bf d}-vector in general.  The deviation from the structure of unitary pairing 
is determined by the balance of energy gain due to (\ref{so15}) and the loss 
of condensation free energy $F_{\rm cond}$.  Although it is not easy to 
compare both effects at arbitrary temperature $0<T<T_{\rm c}$, it becomes 
rather easy at $T\sim T_{\rm c}$, in the so-called GL region.  
The $F_{\rm cond}$ in the ESP state in the GL region, 
is given by the GL free energy 
\begin{equation}
F_{\rm GL}={1\over 2}\left({{\rm d}n\over{\rm d}\epsilon}\right)
\left[-\left(1-{T\over T_{\rm c}}\right)
{\Delta_{\uparrow}^{2}+\Delta_{\downarrow}^{2}\over 2}
+{7\zeta(3)\over16}{\kappa\over 
(\pi k_{\rm B}T_{\rm c})^{2}}
{\Delta_{\uparrow}^{4}+\Delta_{\downarrow}^{4}\over 2}
\right],
\label{so29}
\end{equation}
where $({{\rm d}n/{\rm d}\epsilon})$ is the density of states (DOS) near the 
Fermi level of quasiparticles in the normal state, 
$\kappa\equiv \sum_{\bf k}|{\bf d}_{\bf k}|^{4}/
(\sum_{\bf k}|{\bf d}_{\bf k}|^{2})^{2}$, and 
$\Delta_{{\uparrow}(\downarrow)}$ is the $\uparrow\uparrow$ 
($\downarrow\downarrow$) component of the gap matrix.  
In the unitary state where $\Delta_{\uparrow}=\Delta_{\downarrow}=\Delta$, 
minimizing (\ref{so29}) with respect to $\Delta$, $F_{\rm cond}^{\rm unit}$ 
is given as follows:
\begin{equation}
F_{\rm cond}^{\rm unit}=-{1\over 4}\left({{\rm d}n\over{\rm d}\epsilon}\right)
{8\over 7\zeta(3)}{1\over \kappa}
(\pi k_{\rm B}T_{\rm c})^{2}\left(1-{T\over T_{\rm c}}\right)^{2}.
\label{so30}
\end{equation}
In the GL region, the coupling $g_{\rm d}$ is given as~\cite{Leggett} 
\begin{equation}
g_{\rm d}={\pi\over 8}\mu_{\rm eff}^{2}
\left({{\rm d}n\over{\rm d}\epsilon}\right)^{2}
{8\over 7\zeta(3)}{1\over \kappa}
(\pi k_{\rm B}T_{\rm c})^{2}
\left[\ln(1.14\beta_{\rm c}\epsilon_{\rm c})\right]^{2}
\left(1-{T\over T_{\rm c}}\right).
\label{so31}
\end{equation}
Then, considering the case of Sr$_2$RuO$_4$, we use relation (\ref{spinorbit3a}) 
in 2d and obtain 
the ratio of $g_{\rm so}$ and $|F_{\rm cond}^{\rm unit}|$ as follows: 
\begin{equation}
{g_{\rm so}\over |F_{\rm cond}^{\rm unit}|}=
{ m_{\rm band}\over m^{*}}{(4{\bar g}-1)\over {\bar g}^{2}}
4\pi\mu_{\rm B}^{2}\left({{\rm d}n\over{\rm d}\epsilon}\right)
\left[\ln(1.14\beta_{\rm c}\epsilon_{\rm c})\right]^{2}
\left(1-{T\over T_{\rm c}}\right)^{-1}.
\label{so31a}
\end{equation}

\begin{figure}[ht]
\begin{center}
\rotatebox{0}{\includegraphics[width=0.6\linewidth]{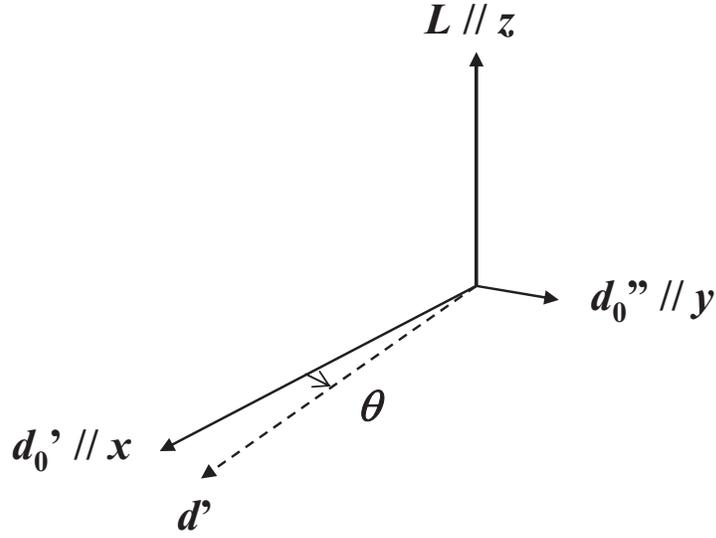}}
\caption{
Structure of {\bf d}-vector. 
}
\label{Fig:1}
\end{center}
\end{figure}

Let us parameterize the gap matrix as 
\begin{equation}
{\hat \Delta}={\Delta_{0}\over\sqrt{1+\eta^{2}}}
\left(
\begin{matrix}
-1-\eta&0\\
0&1-\eta
\end{matrix}
\right).
\label{so32}
\end{equation}
This is equivalent to represent the equilibrium value of {\bf d}-vector as, 
\begin{equation}
d_{0x}={1\over\sqrt{1+\eta^{2}}},\qquad
d_{0y}={\rm i}{\eta\over\sqrt{1+\eta^{2}}}, 
\qquad
d_{0z}=0.
\label{so33}
\end{equation}
This {\bf d}-vector in the equilibrium state is shown in Fig.\ \ref{Fig:1}.  
Substituting expression (\ref{so32}) into (\ref{so29}), after some 
standard calculations, we obtain the loss of condensation free energy 
$\Delta F_{\rm cond}$ as follows:
\begin{equation}
\Delta F_{\rm cond}=|F_{\rm cond}^{\rm unit}|
{4\eta^{2}\over(1+\eta^{2})^{2}}.
\label{so34}
\end{equation}
The energy gain due to the SO coupling (\ref{so15}) is expressed 
in terms of (\ref{so33}) as follows:
\begin{equation}
F_{\rm so}=-g_{\rm so}
{2\eta\over 1+\eta^{2}}.
\label{so35}
\end{equation}
Here we have assumed that the pair angular momentum ${\vec L}$ is 
along the $z$($c$)-axis since we are considering the case of Sr$_2$RuO$_4$.  

Therefore, the total free energy $F(\eta)=F_{\rm so}+
\Delta F_{\rm cond}$ as a function of $\eta$ is given as 
\begin{equation}
F(\eta)=-g_{\rm so}
{2\eta\over 1+\eta^{2}}
+
|F_{\rm cond}^{\rm unit}|
{4\eta^{2}\over(1+\eta^{2})^{2}}.
\label{so36}
\end{equation}
The degree of deviation $\eta$ from the unitary pairing 
is determined by the condition 
$\partial F(\eta)/\partial \eta=0$, which is 
explicitly expressed as 
\begin{equation}
(1-\eta^{2})\left[-g_{\rm so}(1+\eta^{2})
+4|F_{\rm cond}^{\rm unit}|\eta\right]=0.
\label{so37}
\end{equation}
In the case where the deviation from the unitary pairing is small, 
$\eta^{2}\ll 1$ can be neglected compared to unity, so that 
condition (\ref{so37}) is reduced to a simple form
\begin{equation}
\eta={g_{\rm so}\over 4|F_{\rm cond}^{\rm unit}|}.
\label{so38}
\end{equation}
In the case $g_{\rm so}>4|F_{\rm cond}^{\rm unit}|$, condition 
(\ref{so37}) is reduced 
\begin{equation}
\eta=1,
\label{so38a}
\end{equation}
which implies that ${\bf d}_{0}^{\prime}\perp{\bf d}_{0}^{\prime\prime}$; 
non-unitary state is formed.  

The prefactor of $(1-T/T_{\rm c})$ in eq.(\ref{so31a}) is estimated as 
\begin{equation}
{ m_{\rm band}\over m^{*}}{(4{\bar g}-1)\over {\bar g}^{2}}
4\pi\mu_{\rm B}^{2}\left({{\rm d}n\over{\rm d}\epsilon}\right)
\left[\ln(1.14\beta_{\rm c}\epsilon_{\rm c})\right]^{2}
\simeq4.8\times10^{-4}{m_{\rm band}\over m},
\label{so39a}
\end{equation}
where we have used the following relations, 
${\bar g}\simeq1$, $({\rm d}n/{\rm d}\epsilon)=m^{*}/c\pi\hbar^{2}$, 
with $c=6.4$\AA~\cite{Mack}, and we have 
assumed $1.14\beta_{\rm c}\epsilon_{\rm c}\simeq 20$.  
Then, ratio (\ref{so31a}) is estimated to be 
\begin{equation}
{g_{\rm so}\over |F_{\rm cond}^{\rm unit}|}\simeq
4.8\times10^{-4}{m_{\rm band}\over m}\left(1-{T\over T_{\rm c}}\right)^{-1}.
\label{so39}
\end{equation}
Therefore, except for narrow temperature region near $T\simeq T_{\rm c}$, 
the condition $\eta^{2}\ll 1$ holds so that the relation 
(\ref{so38}) is valid.  Since the energy gain due to 
the SO coupling is multiplied by a small factor $\eta$, (\ref{so39}), 
as in eq.(\ref{so35}), in order that the SO interaction predominates over the DD 
interaction, eq.(\ref{so22}), in the low-temperature region, we need another 
mechanism to make the ${\bf d}$ vector within the $ab$-plane, 
as will be discussed in the next section.  

The temperature region where the effect of SO interaction predominates 
over the DD interaction is severely restricted near $T_{\rm c}$ as 
$(1-T/T_{\rm c})<10^{-3}$, taking into account the relations 
eqs.(\ref{so21}), (\ref{so28}), (\ref{so28a}), (\ref{so35}), (\ref{so38}), 
(\ref{so39}), and $m_{\rm band}/m\simeq 2.9$~\cite{Mack}.  Therefore, it is 
technically impossible to observe the non-unitary state by any probe 
considering the broadening of the $T_{\rm c}$ itself.  

\section{Origin of Anisotropy of {\bf d}-Vector - Anisotropy Fields}
The free energy giving anisotropy of {\bf d}-vector due to the magnetic fields 
in ESP state is expressed as~\cite{Leggett}
\begin{equation}
\Delta F_{\rm magn}={1\over 2}\chi_{z}{1-Y(T)\over 1+F_{0}^{\rm a}Y(T)}
({\bf d}\cdot {\bf H})^{2},
\label{An1}
\end{equation}
where $F_{0}^{\rm a}$ is the Fermi liquid parameter for the spin 
susceptibility; 
$\chi_{z}=(1+F_{0}^{a})^{-1}\mu_{\rm B}^{2}({\rm d}n/{\rm d}\epsilon)$.  
In the GL region, this is reduced to 
\begin{eqnarray}
\Delta F_{\rm magn}&=&{1\over \kappa}
{{\displaystyle 1-{T\over T_{\rm c}}}\over 1+F_{0}^{\rm a}}\chi_{z}H^{2}
\nonumber
\\
&=&{1\over \kappa}
{{\displaystyle 1-{T\over T_{\rm c}}}\over (1+F_{0}^{\rm a})^{2}}
\mu_{\rm B}^{2}\left({{\rm d} n\over {\rm d}\epsilon}\right)H^{2}. 
\label{An2} 
\end{eqnarray}
This should be compared with the SO coupling constant 
\begin{equation}
g_{\rm so}={m_{\rm band}\over m^{*}}8{(4{\bar g}-1)\over{\bar g}^{2}}
{\pi\over 8}\mu_{\rm B}^{2}
\left({{\rm d}n\over{\rm d}\epsilon}\right)^{2}
{8\over 7\zeta(3)}{1\over \kappa}
(\pi k_{\rm B}T_{\rm c})^{2}
\left[\ln(1.14\beta_{\rm c}\epsilon_{\rm c})\right]^{2}
\left(1-{T\over T_{\rm c}}\right),
\label{An3}
\end{equation}
where we have used relations (\ref{spinorbit3a}) and (\ref{so31}).  Therefore, 
\begin{equation}
{g_{\rm so}\over \Delta F_{\rm magn}}={1\over H^{2}}
{m_{\rm band}\over m^{*}}(1+F_{0}^{\rm a})^{2}
{8(4{\bar g}-1)\pi^{3}\over 7\zeta(3){\bar g}^{2}}
\left({{\rm d}n\over{\rm d}\epsilon}\right)
(k_{\rm B}T_{\rm c})^{2}
\left[\ln(1.14\beta_{\rm c}\epsilon_{\rm c})\right]^{2}.
\label{An4}
\end{equation}
The actual competition between the effects of the SO coupling and 
the magnetic fields is given by $g_{\rm so}\eta/\Delta F_{\rm magn}$: 
In the region $\eta^{2}\ll 1$, 
\begin{eqnarray}
{g_{\rm so}\eta\over\Delta F_{\rm magn}}
&\simeq&{1\over H^{2}}(1+F_{0}^{a})^{2}{2\pi^{2}\over7\zeta(3)}
\left\{{ m_{\rm band}\over m^{*}}{(4{\bar g}-1)\over {\bar g}^{2}}
4\pi\mu_{\rm B}^{2}\left({{\rm d}n\over{\rm d}\epsilon}\right)
\left[\ln(1.14\beta_{\rm c}\epsilon_{\rm c})\right]^{2}\right\}^{2}
\nonumber
\\
& &\qquad\qquad\qquad\qquad\qquad\qquad
\times{1\over4}\left({k_{\rm B}T_{\rm c}\over\mu_{\rm B}}\right)^{2}
\left(1-{T\over T_{\rm c}}\right)^{-1}\quad\hbox{[gauss$^{-2}]$} 
\label{An5}
\\
&\simeq&3.0\times10^{1}{1\over H^{2}}
(1+F_{0}^{a})^{2}\left({m_{\rm band}\over m}\right)^{2}
T_{\rm c}^{2}
\left(1-{T\over T_{\rm c}}\right)^{-1}\quad\hbox{[gauss$^{-2}]$}, 
\label{An6}
\end{eqnarray}
where we have used the estimation (\ref{so39a}), ${\bar g}\simeq1$, 
$({\rm d}n/{\rm d}\epsilon)=m^{*}/c\pi\hbar^{2}$, 
with $c=6.37\AA$~\cite{Mack}, and we have 
assumed $1.14\beta_{\rm c}\epsilon_{\rm c}\simeq 20$.  
With the use of experimental values of Sr$_2$RuO$_4$, 
$T_{\rm c}\simeq1.5$K, $m_{\rm band}/m\simeq2.9$~\cite{Mack}, 
and $(1+F_{0}^{a})\simeq 0.5$~\cite{Maeno3}, 
\begin{equation}
{g_{\rm so}\eta\over\Delta F_{\rm magn}}
\simeq1.4\times10^{2}{1\over H^{2}}
\left(1-{T\over T_{\rm c}}\right)^{-1}\quad\hbox{[gauss$^{-2}]$}. 
\label{An7}
\end{equation}
Then, the anisotropy field $H_{\rm a}^{{\rm so}(2)}$ due to the two-body SO effect 
is given by the condition 
$g_{\rm so}\eta/\Delta F_{\rm magn}=1$: 
\begin{equation}
H_{\rm a}^{{\rm so}(2)}\simeq1.2\times10\left(1-{T\over T_{\rm c}}\right)^{-1/2}
\quad\hbox{[gauss]}.
\label{An8}
\end{equation}

In the limit of $T\to T_{\rm c}$, $\eta$ approaches 1 so that the 
ratio $g_{\rm so}\eta/\Delta F_{\rm magn}$ is estimated as 
\begin{eqnarray}
{g_{\rm so}\eta\over\Delta F_{\rm magn}}
&\simeq& 0.25\times10^{6}{1\over H^{2}}
(1+F_{0}^{a})^{2}{m_{\rm band}\over m^{*}}T_{\rm c}^{2}
\quad\hbox{[gauss$^{-2}]$}
\label{An9}
\\
&\simeq&4.1\times10^{5}{1\over H^{2}}\quad\hbox{[gauss$^{-2}]$}.
\label{An10}
\end{eqnarray}
Then, $H_{\rm a}^{\rm so(2)}$ is given as 
\begin{equation}
H_{\rm a}^{\rm so(2)}\simeq6.4\times10^{2}\quad\hbox{[gauss]}.
\label{An11}
\end{equation}

The dipole-dipole interaction also gives rise to the anisotropy of 
{\bf d}-vector~\cite{Hasegawa}, as discussed in superfluid $^3$He~\cite{Leggett,LT}  
The favorite direction of {\bf d}-vector is along {\bf L}, relative angular 
momentum of Cooper pairs, i.e., $c$-axis, in contrast to the effect of the SO 
interaction which works to put the {\bf d}-vector in the ab-plane.  
By the analysis similar to the above, using expressions (\ref{so28a}), (\ref{so31}), 
and (\ref{An2}),
\begin{equation}
{(3c/4a\pi)g_{\rm d}\over\Delta F_{\rm magn}}
\simeq 0.39\times10^{4}{1\over H^{2}}
(1+F_{0}^{a})^{2}{m^{*}\over m}T_{\rm c}^{2}
\simeq 3.6\times10^{4}{1\over H^{2}}\quad\hbox{[gauss$^{-2}]$},
\label{An12}
\end{equation}
where, in deriving the second relation, we have used 
$c/a=1.65$, $T_{\rm c}\simeq 1.5$K, $m^{*}/m\simeq 16.0$~\cite{Mack}, and 
$(1+F_{0}^{a})\simeq 0.5$~\cite{Maeno3}.  
Then, the anisotropy filed $H_{\rm a}^{\rm dd}$ is estimated as 
\begin{equation}
H_{\rm a}^{\rm dd}\simeq 1.9\times10^{2}\quad\hbox{[gauss]}.
\label{An13}
\end{equation}
This cannot predominate over $H_{\rm a}^{\rm so(2)}$, eq.(\ref{An11}), in a very 
restricted temperature region near $T_{\rm c}$, 
as discussed in the last paragraph of \S3.  However, in the wide 
range of temperature $0<T<T_{\rm c}$, $H_{\rm a}^{\rm dd}$ predominates over 
$H_{\rm a}^{\rm so(2)}$.  Therefore, the favorite direction of {\bf d}-vector 
is the $c$-axis as far as the SO interaction and the DD interaction of Cooper 
pairs (both of which are two-body effect) are taken into account, 
which is in contradiction to the Knight shift measurements~\cite{Ishida2,Ishida3}, 
as discussed in the fourth paragraph of \S1.  

Another mechanism giving the anisotropy of the {\bf d}-vetor is an atomic 
SO interaction (single-body effect) together with the Hund's rule 
coupling~\cite{Ng,Ogata,Yanase}.  In particular, Yanase and Ogata 
showed~\cite{Yanase}, on the basis of the 3rd order perturbation calculation 
of multiband Hubbard model for the pairing interaction~\cite{Nomura}, 
that the favorite direction of {\bf d}-vector is the $c$-axis and the 
anisotropy field $H_{\rm a}^{\rm so(1)}$ due to the single-body SO interaction 
is of the order of $10^{-2}T_{\rm c}\sim 2\times10^{2}$ gauss.  
Then, adding $H_{\rm a}^{\rm dd}$, the total anisotropy field 
$H_{\rm a}^{\rm total}=H_{\rm a}^{\rm dd}+H_{\rm a}^{\rm so(1)}$ amounts to 
$H_{\rm a}^{\rm total}\sim 4\times10^{2}$ gauss.  
This is also in contradiction to the Knight shift 
measurements~\cite{Ishida2,Ishida3}, which shows that {\bf d}-vector is 
perpendicular to the $c$-axis down to $H=200$ gauss.  

Quite recently, however, it was shown by Yoshioka and the present 
author\cite{Yoshioka} that 
the stable direction of {\bf d}-vector changes from the $c$-axis to the 
$ab$-plane if we take into account the Coulomb repulsion of electrons on 
2p-orbitals at O site\cite{Hoshihara} which cannot be neglected as shown in 
a band structure calculation~\cite{Oguchi}.  Indeed, the anisotropy field 
$H_{\rm a}^{\rm so(1)}$  is estimated as of 
the order of $10^{-2}T_{\rm c}$ by the calculations similar to ref.\citen{Yanase}.  
There is a possibility that this anisotropy field wins $H_{\rm a}^{\rm dd}$, 
ensuring that the favorite direction of {\bf d}-vector is in the $ab$-plane.  
Furthermore, anisotropy of {\bf d}-vector is in the $ab$-plane arises through 
the process breaking the conservation of $z$-component of Cooper pairs by the 
atomic SO interaction~\cite{Yoshioka2}.  It is noted that this process gives 
rise to a weak non-unitary component in the Cooper pair state.  
 
\section{Internal Josephson Oscillations}
It turns out that the so-called internal Josephson effect due to the SO 
coupling is possible.  When the {\bf d}-vector in the equilibrium 
is given by eq.(\ref{so33}), its real and imaginary parts, 
${\bf d}_{0}^{\prime}$ and ${\bf d}_{0}^{\prime\prime}$, are 
\begin{equation}
{\bf d}_{0}^{\prime}={1\over\sqrt{(1+\eta^{2})}}{\hat{\bf x}},
\qquad
{\bf d}_{0}^{\prime\prime}={\eta\over\sqrt{(1+\eta^{2})}}{\hat{\bf y}}, 
\label{so40}
\end{equation}
where ${\hat{\bf x}}$ and ${\hat{\bf y}}$ are the basis vector in the $x$- and 
$y$-directions.  
As the real component of {\bf d}-vector, ${\bf d}_{0}^{\prime}$, 
deviates from equilibrium as shown in Fig.\ \ref{Fig:1}, the {\bf d}-vector is 
\begin{equation}
{\bf d}^{\prime}={\cos\theta\over\sqrt{(1+\eta^{2})}}{\hat{\bf x}}
+\sin\theta\,{\bf d}_{0}^{\prime\prime},
\qquad
{\bf d}^{\prime\prime}={\bf d}_{0}^{\prime\prime}. 
\label{so40a}
\end{equation}
Then, the diagonal components of the gap matrix are given as follows:
\begin{eqnarray}
-d_{x}+{\rm i}d_{y}&=&{1\over\sqrt{1+\eta^{2}}}
(-\cos\theta -\eta+{\rm i}\sin \theta)
\nonumber
\\
&=&{\sqrt{1+2\eta\cos\theta+\eta^{2}}\over\sqrt{1+\eta^{2}}}
{\rm exp}\left[-{\rm i}\tan^{-1}
\left({\sin\theta\over \cos\theta+\eta}\right)\right],
\label{so41}
\end{eqnarray}
and
\begin{eqnarray}
d_{x}+{\rm i}d_{y}=&=&{1\over\sqrt{1+\eta^{2}}}
(\cos\theta -\eta+{\rm i}\sin \theta)
\nonumber
\\
&=&{\sqrt{1-2\eta\cos\theta+\eta^{2}}\over\sqrt{1+\eta^{2}}}
{\rm exp}\left[{\rm i}\tan^{-1}
\left({\sin\theta\over \cos\theta-\eta}\right)\right].
\label{so42}
\end{eqnarray}
Therefore, the phase difference $\Delta \theta$ between $\uparrow\uparrow$- 
and $\downarrow\downarrow$-component of gap matrix is calculated as 
\begin{eqnarray}
\Delta\theta&=&\tan^{-1}\left({\sin\theta\over \cos\theta+\eta}\right)
+\tan^{-1}\left({\sin\theta\over \cos\theta-\eta}\right)
\nonumber
\\
&=&\tan^{-1}\left({\sin 2\theta\over \cos2\theta -\eta^{2}}\right).
\label{so43}
\end{eqnarray}
Up to the ${\cal O}(\eta^{2})$, it is approximately given by $2\theta$:
\begin{equation}
\Delta\theta=2\theta[1+{\cal O}(\eta^{2})].
\label{so44}
\end{equation}
For the small oscillations for which $\theta\ll1$, the gap matrix is given 
up to ${\cal O}(\eta^{2})$ as follows:
\begin{equation}
{\hat\Delta}=\Delta_{0}\left(
\begin{matrix}
(-1-\eta)e^{-{\rm i}\theta}&0\\
0&(1+\eta)e^{{\rm i}\theta}
\end{matrix}
\right).
\label{so45}
\end{equation}

With the use of {\bf d}-vector, eq.(\ref{so40a}), the ``pair-spin" is calculated 
as 
\begin{equation}
({\rm i}{\vec d}\times{\vec d}^{*})_{z}=
2({\vec d}^{\prime}\times{\vec d}^{\prime\prime})_{z}=
2\eta\cos\theta. 
\label{so46}
\end{equation}
Then, the free energy due to SO coupling (\ref{so15}) is 
\begin{equation}
F_{\rm so}(\theta)=-g_{\rm so}(2\eta)\cos\theta,
\label{so47}
\end{equation}
where we have assumed that ${\vec L}={\hat{\bf z}}$ as discussed above.  

The gap structure, eq.(\ref{so45}), and the free energy, eq.(\ref{so46}), have almost 
the same structure as those for the case of dipolar coupling in the A-phase of 
superfluid $^3$He~\cite{Leggett}.  Only difference is that 
$|\Delta_{\uparrow\uparrow}|$ and $|\Delta_{\downarrow\downarrow}|$ are 
slightly different and has weak $\theta$-dependence.  Even though, the equation 
of motion for $\Delta\theta$ and $\Delta N\equiv N_{\uparrow}-N_{\downarrow}$, 
$N_{\uparrow(\downarrow)}$ being the electron number with the spin 
$\uparrow(\downarrow)$, are 
\begin{equation}
{{\rm d}\over {\rm d}t}(N_{\uparrow}-N_{\downarrow})=
-{1\over \hbar}{\partial F_{\rm so}\over\partial(2\theta)}=
-{g_{\rm so}\eta\over \hbar}\sin\theta,
\label{so48}
\end{equation}
and 
\begin{equation}
{{\rm d}\over {\rm d}t}(2\theta)=2\Delta\mu=
{2\mu_{\rm B}^{2}/\hbar\over\chi_{z}}(N_{\uparrow}-N_{\downarrow}).
\label{so49}
\end{equation}
These coupled equations, eqs. (\ref{so48}) and (\ref{so49}), describe the harmonic 
oscillations whose angular eigen frequency is given by 
\begin{eqnarray}
\Omega^{2}&=&{g_{\rm so}(\mu_{\rm B}/\hbar)^{2}\over\chi_{z}}\eta
\\
&=&{g_{\rm so}(\mu_{\rm B}/\hbar)^{2}\over\chi_{z}}
{g_{\rm so}\over 4|F_{\rm cond}|},
\label{so50}
\end{eqnarray}
where expression, eq.(\ref{so38}), has been used for $\eta\ll 1$.  
With the use of eqs.(\ref{so31a}), (\ref{An3}), and 
$\chi_{z}=(1+F_{0}^{a})^{-1}\mu_{\rm B}^{2}({\rm d}n/{\rm d}\epsilon)$, 
$\Omega^{2}$, eq.(\ref{so50}), is expressed as 
\begin{equation}
\Omega^{2}=(1+F_{0}^{a})\left({k_{\rm B}T_{\rm c}\over\mu_{\rm B}}\right)^{2}
{2\pi^{2}\over7\zeta(3)}{1\over \kappa}
\left\{{ m_{\rm band}\over m^{*}}\times{(4{\bar g}-1)\over {\bar g}^{2}}
4\pi\mu_{\rm B}^{2}\left({{\rm d}n\over{\rm d}\epsilon}\right)
\left[\ln(1.14\beta_{\rm c}\epsilon_{\rm c})\right]^{2}\right\}^{2}.
\label{so51a}
\end{equation}
Substituting eq.(\ref{so39a}), we obtain the eigen frequency $\Omega$ as 
\begin{equation}
\Omega\simeq 4.3\times10^{7}\sqrt{(1+F_{0}^{a})/\kappa}\,T_{\rm c}
{m_{\rm band}\over m}\quad{\hbox{[{\rm sec}$^{-1}$]}}.
\label{so51}
\end{equation}

\section{NQR Relaxation Rate due to Internal Josephson Oscillations}
The dynamical uniform spin susceptibility in the A-phase in the ESP manifold 
has been established in the context of superfluid $^3$He by Leggett and Takagi 
as~\cite{LT}:
\begin{equation}
\chi_{z}(\omega)=-{\Omega^{2}\chi_{z}\over 
\omega^{2}-\Omega^{2}+{\rm i}\Gamma\omega},
\label{so52}
\end{equation}
where the damping rate $\Gamma$ is give as  
\begin{equation}
\Gamma=\gamma_{0}\tau\Omega^{2},
\label{so53}
\end{equation}
where $\tau$ is the lifetime of quasiparticles at normal state given as
\begin{equation}
\tau=b{\hbar T_{\rm F}\over k_{\rm B}T^{2}}
=7.6\times10^{-12}b{T_{\rm F}\over T^{2}},
\label{so55}
\end{equation}
where $b$ is a constant of ${\cal O}(1)$.  The coefficient $\gamma_{0}$ in 
eq.(\ref{so53}) is defined as 
\begin{equation}
\gamma_{0}\equiv[1-Y(T)]^{-1}Y(T){\chi_{z}\over
{\displaystyle \mu_{\rm B}^{2}\left({{\rm d}n\over{\rm d}\epsilon}\right)}},
\label{so54}
\end{equation}
where $Y(T)$ is the Yosida function for the $p$-wave ESP state~\cite{Leggett}.  

The NQR/NMR relaxation rate is given by 
\begin{equation}
{1\over T_{1}T}={A\over \mu_{\rm B}^{2}}
\sum_{q<q_{\rm c}}{{\rm Im}\chi_{z}(q,\omega)
\over\omega},
\label{so57}
\end{equation}
where $A$ is a constant arising from the coupling between nuclear and 
electron spin fluctuations.  Imaginary part of $\chi_{z}(\omega)$, 
eq.(\ref{so52}), is 
\begin{equation}
{{\rm Im}\chi_{z}(\omega)\over \omega}=\chi_{z}
{\gamma_{0}\tau\over
{\displaystyle 
\left[\left({\omega\over\Omega}\right)^{2}-1\right]^{2}}+
(\gamma_{0}\omega\tau)^{2}}.
\label{so56}
\end{equation}
Since expression (\ref{so56}) is valid for the wave number smaller than 
inverse of the coherence length, the size of the Cooper pair, the 
cut-off wave number $q_{\rm c}^{*}$ should be set as 
$q_{\rm c}^{*}\sim r(\pi/\xi_{0})$, where $r$ ($<$1) parameterize the 
cut-off size.  

Then, considering the case of Sr$_2$RuO$_4$ where $\omega\simeq1\times10^{7}$ 
and eq.(\ref{so51}), the NQR relaxation rate due to the internal Josephson 
oscillations is given as
\begin{equation}
\left({1\over T_{1}T}\right)_{\rm S(J)}
\simeq {A\over \mu_{\rm B}^{2}}
 \chi_{z}{\pi\over 4}n_{\rm L}cr^{2}\left({a\over \xi_{0}}\right)^{2}
{\gamma_{0}\tau\over 1+(\gamma_{0}\omega\tau)^{2}},
\label{so58}
\end{equation}
where $n_{\rm L}$ is the 3d number density of lattice sites, and 
$n_{2d}$ is the areal number density of quasiparticles.  
Here, the factor $(\omega/\Omega)^{2}$ has been neglected compared to unity, 
because $Omega$, eq.(\ref{so51}), is estimated as $\sim 10^{8}\sim 10\omega$ using 
$(1+F^{a}_{0})\simeq 1/2$~\cite{Maeno3}, $\kappa\sim 1$, $T_{\rm c}\simeq 1.5$, 
and $m_{\rm band}/m\simeq 2.9$~\cite{Mack}.
The ratio of $\xi_{0}$ and the lattice constant $a$ in the plane is 
estimated as in the BCS model:
\begin{equation}
{\xi_{0}\over a}=1.1\times10^{-1}{T_{\rm F}\over T_{\rm c}}.
\label{so59}
\end{equation}
With the use of the experimental values, $\xi_{0}=1050$\AA, 
and $a=3.87$\AA~\cite{Mack}, 
$T_{\rm F}/T_{\rm c}$ is given in turn as
\begin{equation}
{T_{\rm F}\over T_{\rm c}}\simeq 2.5\times10^{3}.
\label{so59a}
\end{equation}
In the cylindrical or 2d model, the DOS is given as follows:
\begin{equation}
\left({{\rm d}n\over{\rm d}\epsilon}\right)\simeq{1\over c}
{n_{2d}\over k_{\rm B}T_{\rm F}}.
\label{so60}
\end{equation}
Then, the relaxation rate, (\ref{so58}), due to the internal Josephson 
effect is expressed as 
\begin{equation}
\left({1\over T_{1}T}\right)_{\rm S(J)}
=6.5\times10{A\over \mu_{\rm B}^{2}}{n_{\rm L}c^{2}\over n_{2d}}
br^{2}\chi_{z}\hbar
\left({{\rm d}n\over{\rm d}\epsilon}\right)\left({T_{\rm c}\over T}\right)^{2}
{\gamma_{0}\over 1+(\gamma_{0}\omega\tau)^{2}}.
\label{so61}
\end{equation}
This expression should be compared with the Korringa relation in the 
normal Fermi liquid state:
\begin{eqnarray}
\left({1\over T_{1}T}\right)_{\rm N}&=&A\sum_{\bf q}{{\rm Im}
\chi_{\rm N}^{\perp}({\bf q},\omega)\over \omega}
\label{so62a}
\\
&\approx&
A\sum_{\bf q}{\rm Im}\left[{\chi_{0}({\bf q},\omega)\over 
1+f_{0}^{a}\chi_{0}({\bf q},\omega)}\right]
\label{so62b}
\\
&\approx&
A{\sum_{\bf q}{\rm Im}\chi_{0}({\bf q},\omega)/\omega\over 
(1+F_{0}^{a})^{2}}
\label{so62c}
\\
&=&A{1\over 4}{\chi_{z}\over \mu_{\rm B}^{2}}
{\hbar{\displaystyle \left({{\rm d}n\over{\rm d}\epsilon}\right)}\over 
1+F_{0}^{\rm a}}{q_{\rm c}\over k_{\rm F}}n_{\rm L}c^{2}a^{2},
\label{so62}
\end{eqnarray}
where $\chi_{\rm N}^{\perp}({\bf q},\omega)$ is the transverse dynamical 
spin susceptibility in the 
normal state, $\chi_{0}({\bf q},\omega)$ is the particle-hole propagator of 
quasiparticles (with $\chi_{0}(0,0)={1\over 2}({\rm d}n/{\rm d}\epsilon)$, 
$k_{\rm F}$ is the Fermi wave number, and 
$q_{\rm c}\simeq 1/2\sqrt{\pi}a$ is the wave number cut-off representing 
the lattice effect.  

In the case of Sr$_2$RuO$_4$, $\chi_{z}\simeq 2.0\times \mu_{\rm B}^{2}
({\rm d}n/{\rm d}\epsilon)$ or $F_{0}^{\rm a}\simeq -0.5$~\cite{Maeno3}, 
$\gamma_{0}$ defined by eq.(\ref{so54}) is approximately given as 
\begin{equation}
\gamma_{0}\simeq 2.0[1-Y(T)]^{-1}Y(T).
\label{so63}
\end{equation}
The parameter $\omega\tau$ in eqs. (\ref{so56}) and (\ref{so61}) is given 
in the present case, $\omega\simeq 1.0\times10^{7}$\ sec$^{-1}$, as 
\begin{equation}
\omega\tau\simeq 7.6\times10^{-5}\times b{T_{\rm F}\over T^{2}}
\simeq1.9\times10^{-1}{b\over T_{\rm c}}\left({T_{\rm c}\over T}\right)^{2}, 
\label{so64}
\end{equation}
where we have used eqs. (\ref{so55}) and (\ref{so59a}).  

The longitudinal relaxation rate of NQR, normalized by that in the normal 
state, eq.(\ref{so62}), is 
\begin{equation}
{(1/T_{1}T)_{\rm S(J)}\over(1/T_{1}T)_{\rm N}}=
{6.5\times10\over (q_{\rm c}/k_{\rm F})a^{2}n_{2d}}br^{2}4(1+F_{0}^{a})
\left({T_{\rm c}\over T}\right)^{2}{\gamma_{0}\over 1+(\gamma_{0}\omega\tau)^{2}}.  
\label{so65}
\end{equation}
The Yosida function $Y(T)$ in $\gamma_{0}$, eq.(\ref{so63}), is estimated in the 
standard manner by assuming that the pairing interaction 
$V_{{\bf k},{\bf k}'}=-V\cos\varphi_{\bf k}\cos\varphi_{{\bf}'}$, 
$\varphi_{\bf k}$ being the azimuth in the {\bf k}-space of $ab$-plane, 
and that the superconducting gap $\Delta_{\rm k}=\Delta\cos\varphi_{\bf k}$ follows 
the weak-coupling gap equation.  Since, in expression (\ref{so65}), there exist 
parameters $b$ and $r$ that are difficult to estimate microscopically, we choose 
them so as to reproduce the observed temperature dependence of NQR relaxation 
rate~\cite{Mukuda}.  
Instead of $b$ and $r$, two independent parameters can be chosen also as 
\begin{equation}
C\equiv {6.5\times10\over (q_{\rm c}/k_{\rm F})a^{2}n_{2d}}br^{2},
\label{so66}
\end{equation}
and 
\begin{equation}
D\equiv 0.39{b\over T_{\rm c}}.
\label{so67}
\end{equation}

\begin{figure}[ht]
\begin{center}
\rotatebox{0}{\includegraphics[width=0.9\linewidth]{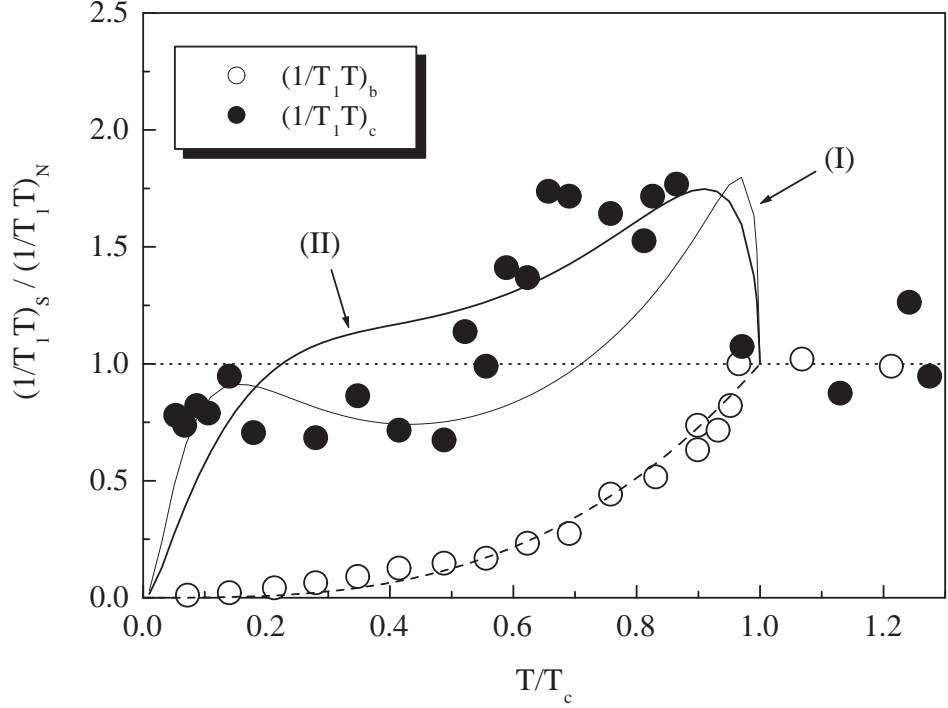}}
\caption{
NQR relaxation rate: Comparison of experiment and theory. Filled (open) circles 
represent data of measurements of NQR relaxation rate due to spin fluctuations 
along the $c$-axis ($b$-axis)~\cite{Mukuda}.  Dashed line represents 
$(1/T_{1}T)_{\rm S(Q)}$ due to the quasiparticles contribution which follows the 
$T^{3}$-law in a wide region $T<T_{\rm c}$.  Solid lines (I) and (II) represent 
$(1/T_{1}T)_{\rm S}=(1/T_{1}T)_{\rm S(J)}+(1/T_{1}T)_{\rm S(Q)}$ for the 
parameter set of eqs.(\ref{so66}) and (\ref{so67}), 
(I) $C=0.2$, $D=0.44$, and (II) $C=0.55$, $D=1.0$, respectively.  
}
\label{Fig:2}
\end{center}
\end{figure}

In Fig.\ \ref{Fig:2}, we show the results of the NQR relaxation rate in the 
superconducting state, 
$(1/T_{1}T)_{\rm S}=(1/T_{1}T)_{\rm S(J)}+(1/T_{1}T)_{\rm S(Q)}$, 
where $(1/T_{1}T)_{\rm S(Q)}$ is the quasiparticles contribution and is 
replaced by experimental values of $(1/T_{1}T)_{b}$, for two sets of parameters, 
(I) $C=0.2$, $D=0.44$, and (II) $C=0.55$, $D=1.0$.  
Agreement of experimental measurements and the theoretical results, based on 
eq.(\ref{so65}), is rather nice, although the theoretical ones include adjustable 
parameters and relatively crude approximations have been done.  To our best knowledge, 
the unusual relaxation rate $(1/T_{1}T)_{c}$ has not yet been explained.  So, our 
theory may be the first one that explains the unusual behavior of NQR relaxation 
rate~\cite{Mukuda}. 

\section{Summary}
We have obtained a formula for the spin-orbit coupling of the Cooper pairs in ESP 
state of Sr$_2$RuO$_4$, giving rise to the internal Josephson oscillations of 
{\bf d}-vector in the $ab$-plane if the stable direction of {\bf d}-vector is in 
the $ab$-plane.  The latter condition is confirmed by a recent theoretical 
finding of Yoshioka and the present author\cite{Yoshioka} 
that the stable direction of {\bf d}-vector 
is in the $ab$-plane on a realistic model of Sr$_2$RuO$_4$ by taking account of 
atomic spin-orbit interaction and the Hund's rule coupling among electrons 
on 4d orbitals at Ru site.  The anomalous temperature dependence of NQR relaxation rate 
$(1/T_{1}T)_{c}$ was explained by the theoretical formula, eq.(\ref{so65}), due to 
the internal Josephson oscillations of {\bf d}-vector in the $ab$-plane that induces 
the oscillations of spin polarization in the direction of $c$-axis.

\newpage
\section*{Appendix}
In this appendix we show how the factor $m_{\rm band}/m^{*}$ appears 
in (\ref{spinorbit1}) on the basis of an extended Ward-Pitaevskii identity.  
Suppose the system is subject to the low rotation $\delta{\vec\Omega}({\bf r})$ 
which is slowly varying with respect to {\bf r}.  
Then the term $-({\bf r}\times{\bf p})\cdot{\vec\Omega}$ is added 
to the Hamiltonian.  Then, the variation of the Hamiltonian is given by 
the term 
\begin{equation}
-\int{\rm d}{\bf r}\psi_{\alpha}^{\dagger}({\bf r})
\left([{\bf r}\times{\bf p}]\cdot\delta{\vec\Omega}({\bf r})\right)
\psi_{\alpha}({\bf r})
=-\int{\rm d}{\bf r}\psi_{\alpha}^{\dagger}({\bf r})
\left([\delta{\vec\Omega}({\bf r})\times{\bf r}]\cdot{\bf p}]\right)
\psi_{\alpha}({\bf r})
\label{A1}
\end{equation}
where ${\bf p}=-{\rm i}\hbar{\vec{\nabla}}$.  By this perturbation, 
we obtain for the variation of the Green function:
\begin{eqnarray}
\delta G&=&-G(p)({\rm i}{\vec{\nabla}}_{p}\times{\bf p})\cdot\delta
{\vec\Omega}G(p+k)
+{{\rm i}\over 2}G(p)G(p+k)
\nonumber
\\
& &\qquad\qquad
\times\int{{\rm d}^{4}q\over(2\pi)^{4}}
\Gamma_{\alpha\beta,\alpha\beta}(p,q;k)G(q)
({\rm i}{\vec{\nabla}}_{q}\times{\bf q})\cdot\delta{\vec\Omega}G(q+k),
\label{A2}
\end{eqnarray}
where $k=({\bf k},0)$ (we assume {\bf k} to be extremely small).  On the other hand, 
the addition of (\ref{A1}) to the Hamiltonian leads to the transformation of 
the momentum in the Green function as 
\begin{equation}
{\bf p}\rightarrow {\bf p}-m_{\rm band}\delta{\vec\Omega}\times
({\rm i}{\vec{\nabla}}_{p}).
\label{A3}
\end{equation}
Hence, 
\begin{equation}
{\delta G\over\delta{\vec\Omega}}=-m_{\rm band}({\rm i}{\vec{\nabla}}_{p})
\times{\partial G\over \partial{\bf p}}, 
\label{A4}
\end{equation}
as ${\bf k}\to 0$.  Consequently, in the limit of $\delta\Omega\to 0$, 
and ${\bf k}\to 0$, we obtain for $G(p)$ describing the quasiparticles: 
\begin{equation}
m_{\rm band}({\rm i}{\vec{\nabla}}_{p})\times{\partial G^{-1}\over \partial{\bf p}}
=-({\rm i}{\vec{\nabla}}_{p}\times{\bf p})+
{{\rm i}\over 2}\int{{\rm d}^{4}q\over(2\pi)^{4}}
\Gamma^{k}_{\alpha\beta,\alpha\beta}(p,q)\{G(q)
({\rm i}{\vec{\nabla}}_{q}\times{\bf q})G(q)\}_{k}.
\label{A5}
\end{equation}
Since the relation 
\begin{equation}
{\partial G^{-1}\over \partial{\bf p}}
=-{{\bf v}\over a}=-{{\bf p}\over m^{*}a}, 
\label{A6}
\end{equation}
holds for the quasiparticles near the Fermi level, 
the relation (\ref{A5}) near the Fermi level is rephrased as 
\begin{equation}
-{m_{\rm band}\over m^{*}a}({\rm i}{\vec{\nabla}}_{p}\times{\bf p})
=-({\rm i}{\vec{\nabla}}_{p}\times{\bf p})+
{{\rm i}\over 2}\int{{\rm d}^{4}q\over(2\pi)^{4}}
\Gamma^{k}_{\alpha\beta,\alpha\beta}(p,q)\{G(q)
({\rm i}{\vec{\nabla}}_{q}\times{\bf q})G(q)\}_{k}.
\label{A7}
\end{equation}
This explains why the vertex correction of spin-orbit coupling is 
given by $m_{\rm band}/m^{*}$, leading to expression (\ref{spinorbit1}) 
after the factor $1/a$ has been cancelled with the renormalization amplitude 
$a$ of quasiparticles.

\section*{Acknowledgements}
The author is grateful to H. Kohno for enlightening discussion on spin-orbit 
interaction associated with relative motion of two electrons, 
K. Ishida for stimulating discussions on Knight shift experiments, 
Y. Kitaoka and H. Mukuda for paying his attention to the present problem, 
Y. Maeno for his continual encouragements, and Y. Yoshioka for informative 
conversations on anisotropy of {\bf d}-vector of Sr$_2$RuO$_4$.  
This work is supported in part by a
Grant-in-Aid for Scientific Research in Priority Area (No. 17071007) and 
for Specially Promoted Research (No.20001004) from the Ministry of Education, 
Culture, Sports, Science and Technology (Japan).


\vskip36pt
\begin{thebibliography}{99}
\bibitem{Maeno1}
Y. Maeno, H. Hashimoto, K. Yoshida, S. Nishizaki, T. Fujita, J. G. Bednorz, 
and F. Lichtenberg: Nature {\bf 372} (1994) 532.
\bibitem{Ishida1}
K. Ishida, H. Mukuda, Y. Kitaoka, K. Asayama, Z. Q. Mao, Y. Mori and 
Y. Maeno: Nature {\bf 396} (1998) 658.
\bibitem{Leggett}
A. J. Leggett, Rev. Mod. Phys. {\bf 47} (1975) 331.
\bibitem{Rice}
T. M. Rice and M. Sigrist: J. Phys.: Condens. Matter {\bf 7} (1995) L643.
\bibitem{Tou1}
H. Tou, Y. Kitaoka, K. Asayama, N. Kimura, Y. \=Onuki, E. Yamamoto, 
and K. Maezawa: Phys. Rev. Lett. {\bf 77} (1996) 1374.
\bibitem{Tou2}
H. Tou, Y. Kitaoka, K. Ishida, K. Asayama, N. Kimura, Y. \=Onuki, E. Yamamoto, 
Y. Haga, and K. Maezawa: Phys. Rev. Lett. {\bf 80} (1998) 3129.
\bibitem{Yotsuhashi}
S. Yotsuhashi, K. Miyake and H. Kusunose: 
Physica B {\bf 312-313} (2002) 100. 
\bibitem{Ishida2}
H. Murakawa, K. Ishida, K. Kitagawa, Z. Q. Mao, and Y. Maeno: 
Phys. Rev. Lett. {\bf 93} (2004), 167004.
\bibitem{Ishida3}
H. Murakawa, K. Ishida, K. Kitagawa, H. Ikeda, Z. Q. Mao, and Y. Maeno: 
J. Phys. Soc. Jpn. {\bf 76} (2007) 024716. 
\bibitem{Hoshihara}
K. Hoshihara and K. Miyake: J. Phys. Soc. Jpn. {\bf 74} (2005) 2679.
\bibitem{Yoshioka}
Y. Yoshioka and K. Miyake: J. Phys. Soc. Jpn. {\bf 78} (2009) 074701.
\bibitem{Mukuda}
H. Mukuda, K. Ishida, Y. Kitaoka, K. Miyake, Z. Q. Mao, Y. Mori and 
Y. Maeno: Phys. Rev. B {\bf 65} (2002) 132507.
\bibitem{Ishida4}
K. Ishida, H. Mukuda, Y. Kitaoka, Z. Q. Mao, Y. Mori, and Y. Maeno: 
Phys. Rev. Lett. {\bf 84} (2000) 5387.
\bibitem{AGD}
A. A. Abrikosov, L. P. Gorkov, I. E. Dzyaloshinskii: 
{\it Method of Quantum Field Theory in Statistical Physics}, 
2nd ed. (Pergamon Press, Oxford, 1965) \S19.1.
\bibitem{Leggett1964}
A. J. Leggett: Phys. Rev. {\bf 140} (1965) A1869.
\bibitem{Hasegawa}
Y. Hasegawa: J. Phys. Soc. Jpn. {\bf 72} (2003) 2456. 
\bibitem{Mack}
A. P. Mackenzie, S. R. Julian, A. J. Diver, G. J. McMullan, M. P. Ray, 
G. G. Lonzarich, Y. Maeno, S. Nishizaki and T. Fujita, 
Phys. Rev. Lett. {\bf 76} (1996) 3786. 
\bibitem{Maeno3}
Y. Maeno, K. Yoshida, H. Hashimoto, S. Nishizaki, S. Ikeda, M.
Nohara, T. Fujita, A. P. Mackenzie, N. E. Hussey, J. G. Bednorz, and 
F. Lichtenberg: J. Phys. Soc. Jpn. {\bf 66} (1997) 1405.
\bibitem{LT}
A. J. Leggett and S. Takagi: Ann. Phys. {\bf 106} (1977) 79.
\bibitem{Ng}
K. K. Ng and M. Sigrist: Europhys. Lett. {\bf 49} (2000) 473.
\bibitem{Ogata}
M. Ogata: J. Phys. Chem. Solids {\bf 63} (2002) 1329.
\bibitem{Yanase}
Y. Yanase and M. Ogata: J. Phys. Soc. Jpn. {\bf 72} (2003) 673.
\bibitem{Nomura}
T. Nomura and K. Yamada: J. Phys. Soc. Jpn. {\bf 71} (2002) 1993.
\bibitem{Oguchi}
T. Oguchi: Phys. Rev B {\bf 51} (1995) 1385.
\bibitem{Yoshioka2}
Y. Yoshioka: private communications. 
\end{thebibliography}
\end{document}